\documentclass[sigconf]{acmart}
\usepackage[T1]{fontenc}
\usepackage{hyperref}
\usepackage{ragged2e}
\usepackage{tikz}
\usepackage{booktabs}
\usepackage{amsmath}
\usepackage{filecontents}
\usepackage{natbib}
\usepackage{xspace}
\usepackage{subfigure}

\usepackage{balance}
\usepackage{url}

\usepackage{breakurl}

\newcommand{\gens}{Stratosphere\xspace}
\newcommand{\todo}[1]{}
\renewcommand{\todo}[1]{{\color{red} TODO: {#1}}}

\date{}

\title{Stratosphere: Finding Vulnerable Cloud Storage Buckets}

\author{Jack Cable}
\affiliation{%
  \institution{Stanford University}
    \city{}
   \state{}
   \country{}
}
\authornote{The first three authors made substantial contributions.}

\author{Drew Gregory}
\affiliation{%
  \institution{Stanford University}
    \city{}
   \state{}
   \country{}
}
\authornotemark[1]

\author{Liz Izhikevich}
\affiliation{%
  \institution{Stanford University}
    \city{}
   \state{}
   \country{}
}
\authornotemark[1]

\author{Zakir Durumeric}
\affiliation{%
  \institution{Stanford University}
      \city{}
   \state{}
   \country{}
}

\SetExtraKerning{encoding=*,font=*}{\textemdash={120,120}}

\copyrightyear{2021} 
\acmYear{2021} 
\setcopyright{acmlicensed}\acmConference[RAID '21]{24th International Symposium on Research in Attacks, Intrusions and Defenses}{October 6--8, 2021}{San Sebastian, Spain}
\acmBooktitle{24th International Symposium on Research in Attacks, Intrusions and Defenses (RAID '21), October 6--8, 2021, San Sebastian, Spain}
\acmPrice{15.00}
\acmDOI{10.1145/3471621.3473500}
\acmISBN{978-1-4503-9058-3/21/10}

\begin{document}
\begin{abstract}

Misconfigured cloud storage buckets have leaked hundreds of millions of medical, voter, and customer records. These breaches are due to a combination of easily-guessable bucket names and error-prone security configurations, which, together, allow attackers to easily guess and access sensitive data. In this work, we investigate the security of buckets, finding that prior studies have largely underestimated cloud insecurity by focusing on simple, easy-to-guess names. 
By leveraging prior work in the password analysis space, we introduce \gens, a system that learns how buckets are named in practice in order to efficiently guess the names of vulnerable buckets. 
Using \gens, we find wide-spread exploitation of buckets and vulnerable configurations continuing to increase over the years. 
We conclude with recommendations for operators, researchers, and cloud providers.


\end{abstract}
\maketitle

\section{Introduction}

Cloud storage services like Amazon Simple Storage Service (S3), Google Cloud Storage, and Alibaba Cloud Object Storage Service have grown tremendously in popularity over the past 15~years. In 2021, Amazon alone reported that they host more than 100~trillion files in their S3 service~\cite{awsTrillion}. Storage services help developers launch new products by eliminating burdensome administrative tasks like file replication and backup. However, they have also introduced a new attack vector for many companies. Today's services have increasingly complex configuration options and operators regularly misconfigure services in vulnerable ways. Indeed, over the past 18~months, a purported 80\% of companies have experienced a cloud-related data breach~\cite{cloudBreach}. Many of these breaches have been catastrophic, resulting in hundreds of millions of leaked voter registration records~\cite{voterBreach}, login credentials to classified systems~\cite{blackboxupguard}, sensitive medical records~\cite{medical}, and customer PII~\cite{deahl_2017}.


Unfortunately, the cloud storage ecosystem has remained relatively out of view to security researchers, because, unlike conventional servers that are accessible by IP address, most storage buckets are accessible only through their customer-provided name. In the case of Amazon, Google, and Alibaba's storage offerings, buckets can be 3--64~characters long, supporting roughly $10^{101}$~names---$10^{62}$~times larger than the IPv6 search space. This not only limits researchers' ability to understand how and why buckets are left unsecured, but precludes notifying affected organizations. As we discuss in Section~\ref{sec:security}, attackers are scanning for vulnerable buckets and our honeypots see unsolicited traffic within 24~hours whereas it appears to take AWS months (and, in one case, \emph{years}) to notify users of vulnerable storage configurations.  


In this paper, we introduce \gens, a system that learns how humans name storage buckets and efficiently guesses the names of public buckets that leak sensitive data. \gens is not the first system proposed for finding insecure buckets, and we start our study with an analysis of prior work. We find that state-of-the-art (e.g.,~\cite{Continella}) scanners have high hit-rates but only find buckets with short, simple, and easy-to-guess names. These names are not representative of real-world buckets (as found through multiple passive DNS sources), constraining them to underestimating global vulnerability by 500\%. We analyze the buckets found through passive DNS sources and find that while names are more complex than those generated by existing scanners, they are also structured, and, in many cases, predictable.

We identify parallels between the creation of passwords and bucket names---both settings in which human language is used to construct strings to protect data---and build upon the rich literature of password analysis/cracking techniques to architect a system that can predict otherwise hard-to-guess bucket names. \gens learns real-world naming patterns from bucket names leaked both online (e.g., in GitHub repositories) and in passive DNS sinks to learn how buckets are named in practice, and it can find 2.7~times more public buckets, 5.8~times more misconfigured buckets (i.e., full write and delete access) and 5.3~times more buckets that host potentially sensitive documents compared to prior techniques~\cite{Continella}. Furthermore, \gens finds bucket names on average 4~orders of magnitude more complex than prior work---inline with real-world buckets we observe in passive DNS datasets.

We analyze the 2.1M buckets we find, including those that \gens uncovers across Amazon S3, Google Cloud Storage, and Alibaba Object Storage.
Using American Express EarlyBird~\cite{earlybird}, we identify sensitive data being hosted in 10.6\% of public buckets, including a bucket belonging to a Department of Defense contractor. We notify organizations of their exposure and detail the results of our disclosure process.

Our improved perspective also allows us to unearth broader ecosystem patterns. We find that AWS S3 exhibits the worst---and continually worsening---security problems: buckets updated on AWS in the past year are on average 4~times more likely to be vulnerable than buckets updated in the last 10~years. In the worst case, 5\% of \emph{private} AWS buckets with readable permissions allow for their permissions to be changed by any unauthenticated user, showing that misconfigurations remain a pressing problem in identified buckets. We further observe evidence that 3\% of all vulnerable buckets have {already been exploited}.



Our results highlight that cloud storage buckets continue to be misconfigured in 2021, but that the security community can help operators secure their cloud presence. By releasing \gens as an open-source tool, we hope to enable operators and researchers to more accurately understand and improve cloud security. 

\section{Related Work}

Security research has benefited from systems that dramatically increase the performance of Internet data collection and analysis. ZMap~\cite{durumeric2013zmap} and Masscan~\cite{graham2014masscan} have pushed Internet-wide scanning to the theoretical limit, allowing for the discovery of cryptographic weaknesses~\cite{heninger2012mining,hastings2016weak} and uncovering of real-world attacks~\cite{rossow2014amplification,marczak2014governments}.
Snort~\cite{snort} and Bro~\cite{paxson1999bro} have made packet parsing a  lightweight and modular operation, allowing for the subversion of botnets~\cite{botminer,Gu2008BotSnifferDB} and detection of malware infections in real time~\cite{gu2007bothunter}.

We build on two bodies of literature to create a system that approaches the theoretical limit of cloud storage scanning to efficiently find vulnerable cloud storage buckets.
To find valid bucket names used in cloud storage and evaluate the guessability of bucket names, our work leverages  password cracking~\cite{CFG,Cracking} and password-strength checking~\cite{wheeler2016zxcvbn}  techniques, which predict and characterize how humans use language to construct strings to protect their data. Furthermore, we leverage a sequence of studies that focus on efficiently scanning the IPv6 address space by developing target generation algorithms (TGAs) that draw upon sets of existing known IPv6 addresses to infer additional addresses~\cite{ullrich2015reconnaissance,foremski2016entropy,murdock2017target,6tree}. For example, Entropy/IP~\cite{foremski2016entropy} uses a Bayesian Network to find common IPv6 address patterns in order to generate new IPv6 addresses that exhibit a similar address structure. 

Most similar to the security problems we uncover in our work is a study by Continella et~al.~\cite{Continella}, who found that misconfigured buckets could lead to code injection and defacement of websites that use S3 to load web resources. We compare our results to Continella when appropriate. Beyond identifying buckets, past work has investigated security vulnerabilities in other cloud infrastructure~\cite{Ristenpart:2009:HYG:1653662.1653687, EBS, somorovsky2011all,izhikevich2018building}, such as Ristenpart et al., who mapped Amazon's internal IP address space in order to exploit VM co-residency and Somorovsky et al., who found that Amazon control interfaces were vulnerable to XSS attacks.
Short blog posts~\cite{s3honey,s3honey2} and videos~\cite{eroBuckets} discuss the process of setting up and evaluating cloud storage honeypots on AWS S3\@.

\section{Evaluating Cloud Storage Scanners}
\label{sec:motivation}

\begin{table*}[h]
\small
\centering
\begin{tabular}{l l l rrrrrrr }
\toprule
Type &  Source & Collection & Candidates & Public & Total & Sensitive & Misconfigured & \%Valid & \% Unique  \\
& & Time & & & & & & & Valid \\

\midrule
Passive DNS & Farsight  & 7~years &  1,254,682 & 95,648 & 773,595 & 3.3K &  4.1K & 61\%    & 55\%  \\
&VirusTotal  & 8~years & 889,176 & 16,940 & 91,414 & 2.3K & 1.1K  &10\%    & 73\%   \\
&Zetalytics & 6 years & 83,104  & 71     & 1,892 & 1 &    6   &2\%    & 68\%   \smallskip \\ 

Search & Bing & >10~years & 36,093             & 4,180            & 16,998     & 1.1K &   194      &47\%     &     52\%  \\
&GitHub & 5 years & 10,864 & 850        &4,444 &  115 &    37   &41\%     & 24\%   \smallskip  \\

Repository   &Grayhat &2~years  &  115,269 & 37,376        & 42,074      & 2.0K & 2.2K & 37\%    & 31\% \smallskip \\

Scanner &Continella  & 1~month & 6,919,902 & 16,199          & 125,712  & 208 &    635    & 4.6\%   & 21\%  \\
&Random& 1~month &  10,523,161   & 5,002         & 77,985   & 69 &  281   & 0.1\%  & 2.0\% \\
\midrule
New & LSTM & 1.5 months & 16,310,302  & 13,989 & 259,352 & 219 & 1.0K & 1.6\% & 49\% \\
New & Token PCFG & 1.5 months & 14,706,908 &  13,334 & 286,610 & 300 & 1.5K & 2.0\% & 50\% \\
New & Character PCFG & 1.5 months & 16,913,529 & 9,705 & 185,978 & 107 & 543 & 1.1\% & 20\%  \\
New & Token Bigrams & 1.5 months & 10,302,097 & 3,022 & 65,864 & 66 & 327 & 0.64\% & 10\% \\
New & Character 5-Grams & 1.5 months & 40,383,614 & 18,396 & 215,525 & 110 & 528 & 0.53\% & 36\%  \\
\bottomrule
\end{tabular}
\caption{\textbf{Data Sources}---\textnormal{We extract buckets from different categories of data sources to empirically analyze the cloud storage space. At least 50\% of valid buckets  extracted from passive data sources are unique across all sources, thereby highlighting the need to combine many different data sources. Buckets collected by our active scanners are completely unique compared to all passive and search data sources.}
}
\label{table:data_breakdown}
\end{table*}

Vulnerable cloud storage buckets can be found using active scanning and passive sources (e.g., passive DNS).
While buckets found in passive sources are already exposed to the public, passive DNS is a privileged perspective that only uncovers a small sample of all buckets. Furthermore, passive DNS will likely become less effective in the future as DNS-over-HTTPS and DNS-over-TLS become ubiquitously deployed. Efficient cloud storage scanning allows for quickly discovering a greater number of buckets, such as older vulnerable buckets that do not show up in recent Internet traffic.
It is imperative that the security community can not only better understand how and why buckets are left unsecured, but can also quickly notify affected organizations en masse, as cloud providers are inexplicably slow at notifying their customers. We deploy 35 vulnerable AWS storage buckets  with varying name complexity (Appendix~\ref{app:honey_meths}) and receive unsolicited traffic within the first 24~hours. However, it takes AWS \emph{4~months} to notify us that our buckets are vulnerable. 
\looseness=-1

Several groups have proposed solutions for guessing the names of storage buckets. 
The current state-of-the-art bucket generator is a scanner developed by Continella et~al.~\cite{Continella}, which guesses S3 buckets by generating random 3--4 character sequences and performing random operations of removing a character, duplicating a character, or concatenating a new word from a corpus.

There also exist several open source scanners that generate candidate names based on a user-provided word list and template patterns, including Slurp~\cite{slurp}, s3enum~\cite{s3enum}, S3Scanner~\cite{S3Scanner}, and BucketStream~\cite{bucket-stream}. 
These open source scanners are ``target generators''~\cite{s3-mining,slurp,_willis_2013}---they require a target word to search for candidates, such as an organization name---and are not inherently built to scan globally. s3enum~\cite{s3enum} no longer works as AWS has removed the side-channel it relies upon. 
Lastly, there are two public search engines---Grayhat Warfare~\cite{grayhatwarfare} and Public Cloud Storage Search~\cite{pcloudSearch}, the former being most well known. There is little public information about how these services uncover buckets. 


To understand the limits of existing work, we evaluate Continella et~al. and Grayhat Warfare against a ground truth sample of real-world buckets found in passive DNS data sources. We show that while state-of-the-art active scanning methods are efficient at finding valid bucket names, they produce extremely short names and are biased against finding vulnerable buckets. 
We use our understanding of why and how prior work falls short of finding real-world bucket names to build a system tailored towards efficiently finding vulnerable buckets (Section~\ref{sec:generators}).


\subsection{Evaluation Methodology}
\label{sub:sec:gtruth}

One approach for evaluating scanning algorithms would be to simply calculate the hit-rate of found buckets compared to generated candidate names. However, this na\"ive metric does not accurately capture whether generated bucket names are characteristic of real-world buckets where security problems are likely to occur. Furthermore, it does not allow us to evaluate black-box solutions like Grayhat Warfare where we only see found buckets and not any failed lookup attempts. Overall hit-rate also fails to determine whether algorithms are efficient at identifying publicly accessible and ``vulnerable'' buckets. As such, we further evaluate solutions based on publicly accessible hit rate and vulnerability hit rate.

\vspace{3pt}
\noindent \textbf{Definition of Vulnerable.}
We define buckets to be \emph{vulnerable} if they have publicly accessible sensitive content or have misconfigured permissions. We use American Express EarlyBird~\cite{earlybird}, a tool that supplies 60~common sensitive filename patterns like private keys and database dumps to detect whether a bucket exposes sensitive data. We further analyze all publicly-available bucket Access Control Lists (ACLs) and classify misconfigured buckets as those with an ACL that allows public write, public delete, or public modification of the ACL itself. 

\vspace{3pt}
\noindent \textbf{Passive Data Sources.}
To investigate whether existing tools can generate real-world bucket names as well as find vulnerable buckets, we compare generated names against a sample of names found in passive DNS data sources. Amazon Web Services (AWS), Google Cloud Platform (GCP), and Alibaba storage buckets are accessible as cloud-specific subdomains (e.g., \texttt{mybucket.s3.amazonaws.com}). Since most DNS queries are made in cleartext, bucket names are sometimes inadvertently exposed to passive DNS sinks. We collect 2.2M~candidate bucket names from three well-known passive DNS sources: Farsight~\cite{farsight}, VirusTotal~\cite{virustotal}, and Zetalytics~\cite{zetalytics} between October 2019 and June 2020. We additionally search for buckets through Bing via the Azure Cognitive Services API~\cite{cogserv} using the query \texttt{site:bucket\_hostname} and collect 11K~candidate buckets that are exposed in public GitHub repositories by querying the GitHub BigQuery dataset~\cite{githubbq} for files that contain AWS, GCP, and Alibaba bucket URLs.

To confirm that a bucket exists for each name we observe, we perform an HTTP GET request to each candidate bucket using ZGrab~\cite{zgrab2}. We decode response status codes: 404 as nonexistent, 403 as private, and 200 as public. In Alibaba Cloud, an unlistable bucket can return a 200 status code if it is a bucket website, so we consider an Alibaba bucket public if it also allows listing its contents in addition to a 200 response. In total, we uncover 532,284~\emph{unique} valid buckets from passive DNS and search data sources. We do not make any argument that the passive data sources are complete---however, they provide a best-effort approximation of frequently accessed real-world buckets.

\vspace{3pt}
\noindent \textbf{Ethical Considerations.} 
We follow the best practices set forth by Durumeric
et~al.~\cite{durumeric2013zmap} to minimize impact of our research. We received three inquiries about our scans, but no party requested to opt out. In line with prior studies~\cite{springall2016ftp, Continella}, we analyze file metadata (e.g., filename and extension), but never directly view or download files except for the index page of public Alibaba website buckets. In the course of our manual analysis, when we suspect exposed sensitive files based on filenames, we make a best effort to disclose to the bucket owner and/or cloud provider (Appendix~\ref{app:disclose}).


\begin{figure*}[h]
 \centering
\subfigure[Name Length]{%
  \includegraphics[width=0.32\linewidth]{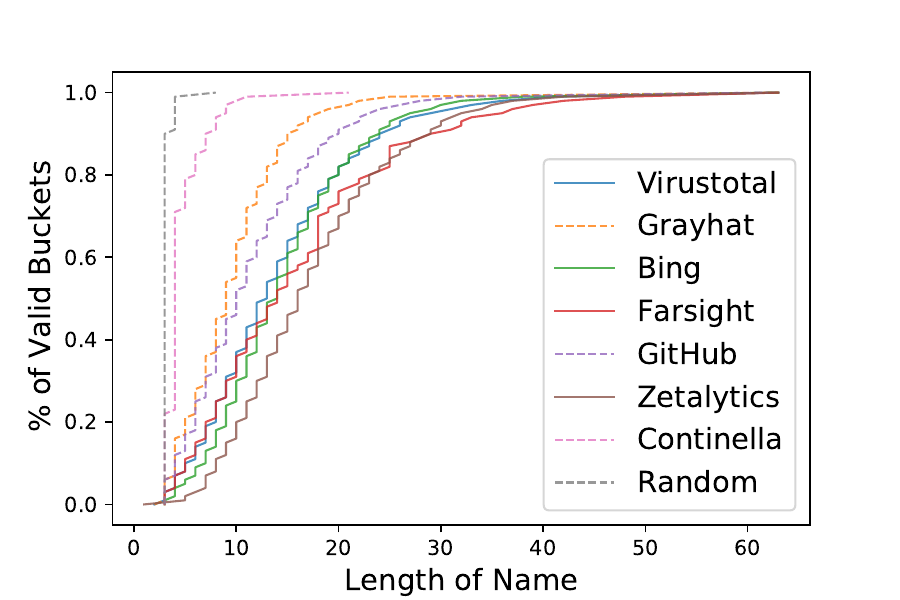}
  \label{fig:gt_length}
  }%
\subfigure[Shannon Entropy]{%

	\includegraphics[width=0.32\linewidth]{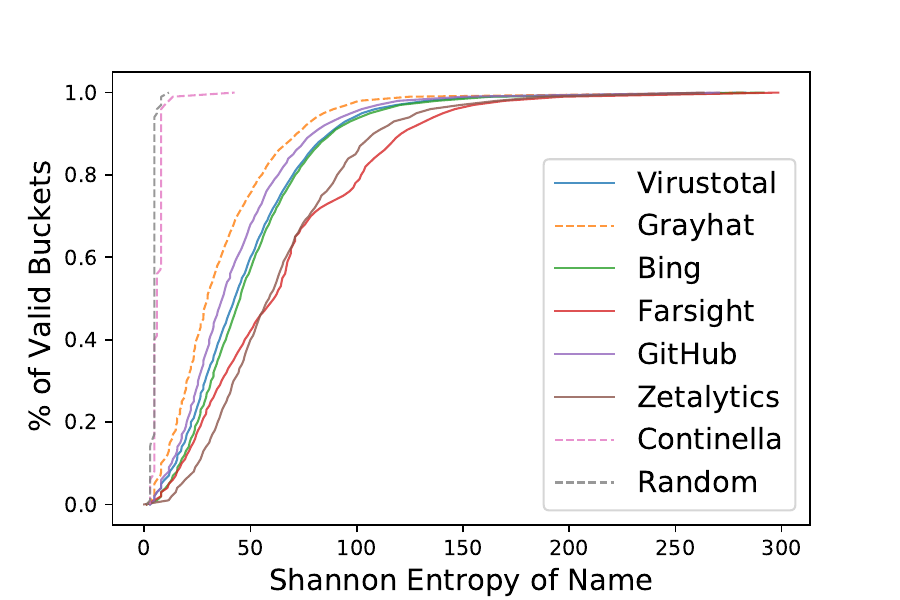}
	\label{fig:gt_shannon}
	  }%
\subfigure[$\log_{10}(Guessability)$ ]{
  \centering
	\includegraphics[width=0.32\linewidth]{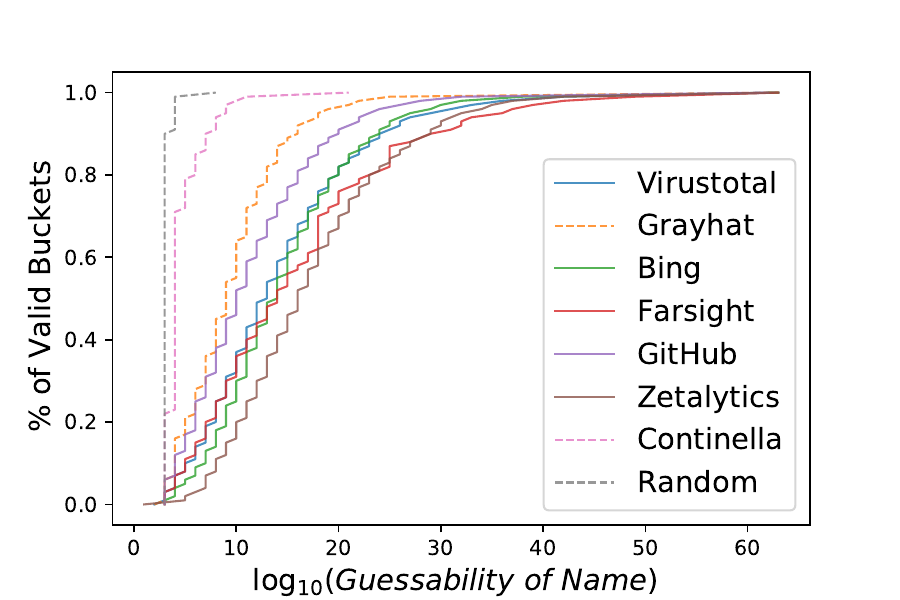}
	\label{fig:gt_name_complexity}
	  }%
	\caption{\textbf{Guessability of Bucket Names}---%
	\textnormal{Bucket names found in Internet traffic repositories tend to be harder to guess (i.e., greater length, Shannon Entropy, and estimated number of guesses required) compared to bucket names found in black-box repositories (e.g., Grayhat) and scanners (e.g., Continella). }
	}

\label{fig:gt_guesses}
\end{figure*}

\subsection{Evaluating State-of-the-Art Scanners}
\label{sub:sec:existing_scanners}

To evaluate the state-of-the-art scanner designed by Continella et~al.~\cite{Continella}, we implement the described algorithm, which guesses buckets by generating random 3--4~character sequences and performing random operations of removing a character, duplicating a character, or concatenating a new word from a corpus.
We note that since Continella et~al. omit all hyper-parameter values used in their algorithm, we assign equal probabilities to the algorithm's parameter of removing a word, concatenating a word, or stopping. Furthermore, we use a naive scanner---as a control experiment---that scans for random alphanumeric combinations between the lengths of 3 and 64. We also evaluate all publicly available buckets collected by the popular cloud storage collection repository, Grayhat Warfare~\cite{grayhatwarfare}, which does not reveal their method of scanning and keeps a majority of found buckets private behind a paywall. 

We run the Continella and Random scanner for the month of December 2020 and collect buckets from Grayhat in December. Using the existing scanning solutions, we discover an additional unique 189K~buckets of which 35K are publicly accessible (Table~\ref{table:data_breakdown}).
Continella finds 540 valid public~buckets per day, which is an order of magnitude faster than passive sources.
Continella achieves an average hitrate of 4.6\%, primarily because over 99\% of valid names guessed by Continella are 3 or 4~characters long, effectively focusing on only a small high-density portion of the name space. However, these short valid bucket names are not representative of real-world buckets: 98\% of valid buckets in passive data sources are longer than 4~characters. The median length, average length, and median Shannon Entropy of Continella generated names is half that of all names found in passive data sources (Figure~\ref{fig:gt_guesses}). The names found by Grayhat are also on average an order of magnitude shorter than passive data sources. 



Furthermore, Continella finds 15~times and 6.4~times fewer sensitive and misconfigured buckets than found in passive datasources, which is disproportionately lower when compared to the total number of buckets found.
We hypothesize that there may exist a correlation between vulnerable buckets and bucket names that are ``harder to guess,'' which may be contributing to Continella's bias against finding vulnerable buckets. To properly quantify the guessability (i.e., complexity) of bucket names, we use  zxcvbn~\cite{wheeler2016zxcvbn}, a popular password analysis tool that provides a fast and low-cost algorithm to match occurrences of words from large corpuses to parts of a string and estimates the minimum number of guesses an attacker can successfully guess a name via a dictionary-driven attack. Zxcvbn accounts for bucket names that might appear long and high in entropy due to containing a domain name or a common word which would otherwise be quickly found in a corpus.
We extend zxcvbn's corpus of known words by adding a well-known dictionary of 466K~common words~\cite{scowl}, domain names from the top 1~million websites~\cite{cisco_umbrella}, common file extensions~\cite{file_extensions}, common technology names (e.g., postgres, sftp) \cite{tech_terms}, and symbols. 

\begin{figure}[h]
    \centering
    \includegraphics[width=\linewidth]{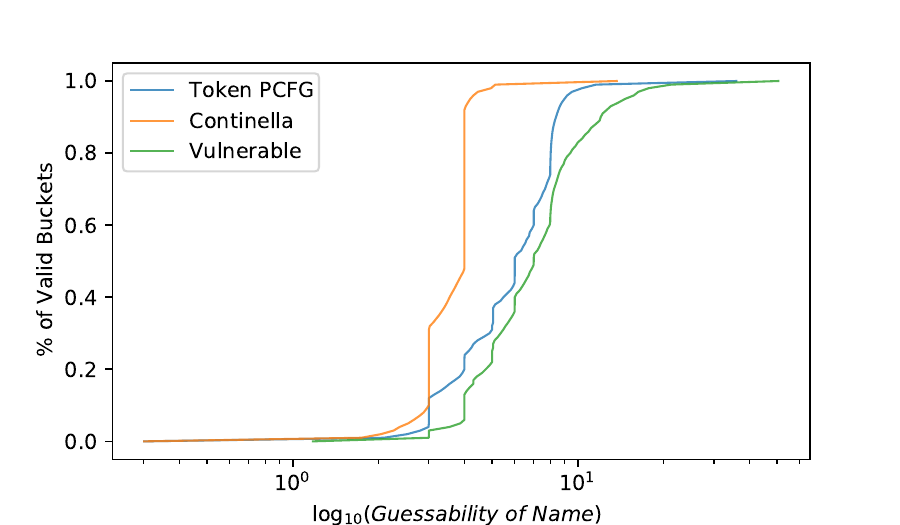}
    \caption{\textbf{Vulnerable Naming Patterns}---\textnormal{Vulnerable bucket configurations have more complex names than prior work, Continella, is able to find. Our best-performing generator, Token-PCFG (Section~\ref{sec:generators}), is more capable of finding bucket names more similar to those exhibited by vulnerable buckets.} }
    \label{fig:vuln_guess}
\end{figure}

We find that vulnerable public buckets are twice as likely to have a harder-to-guess name (per the zxcvbn guessability heuristic) than non-vulnerable buckets. Public buckets are also 1.4x~more likely to be harder to guess than private buckets.
Thus, it is not surprising that Continella is biased against finding vulnerable buckets; the average bucket name found by Continella is four orders of magnitude less complex than the average vulnerable bucket name found in our passive sources (Figure~\ref{fig:vuln_guess}).
If Continella was capable of finding bucket names following the same name-complexity distribution as the buckets found by Farsight, Continella would find an estimated 3.2~times more buckets with sensitive files and 1.6~times more total vulnerable buckets while generating the same number of candidates.

\subsection{Bucket Naming Patterns}

Existing scanners that attempt to exhaustively scan all buckets primarily rely on random permutations of characters and words to find valid bucket names. Consequently, they primarily only find 3--4 character bucket names that are not representative of real-world bucket names and are less likely to be vulnerable.
In this section, we dissect how humans name buckets in practice in order to understand if naming patterns exist and can be leveraged to better predict real-world bucket names. 
We use the popular password analysis tool, zxcvbn (Section~\ref{sub:sec:existing_scanners}), which matches occurrences of words from corpuses to parts of a string. We choose zxcvbn for its efficiency as well as its ability to find ``l33t'' naming conventions, which are modified spellings using numbers and symbols instead of letters. We extend zxcvbn's corpus of known words as described previously in Section~\ref{sub:sec:existing_scanners}.

We run zxcvbn against all public and private buckets found in our least-biased sources: five passive DNS and search sources. We define any part of a bucket name matched to a corpus as a ``corpus'' token and any part not matched as a ``random'' token. For example, the bucket names ``dogs'' and ``dogf'' would be de-constructed as \texttt{corpus} and \texttt{corpus+rand}, respectively. 
Our results show that a password analysis tool is successful at finding popular naming conventions and that the ordering of tokens is predictable. 

\begin{figure}[h]
    \centering
    \includegraphics[width=\linewidth]{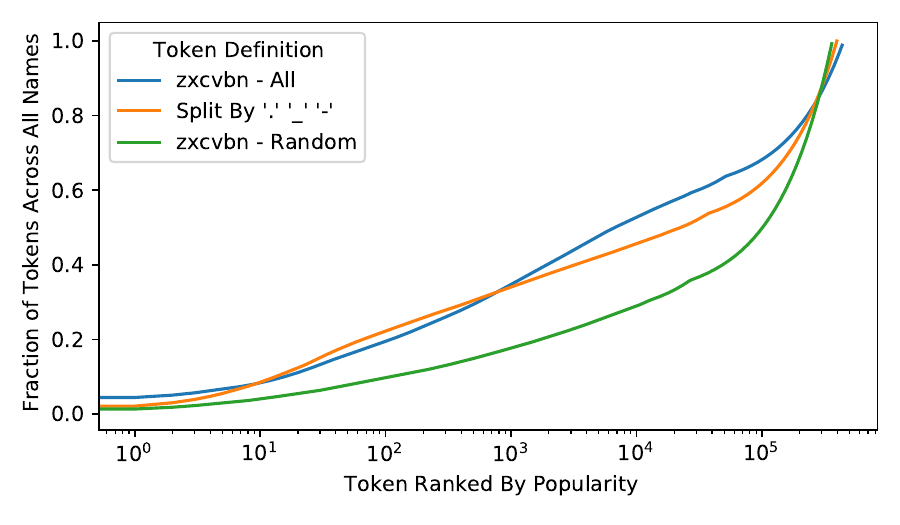}
    \caption{\textbf{Token Popularity}---\textnormal{Tokens in bucket names are often re-used, with the top~100 tokens appearing in 25\% of all names. Random tokens are much more diffuse; the top~100 only appear in 9.8\% of names.}
    }
    \label{fig:token_dist}
\end{figure}

\vspace{5pt}
\noindent \textbf{Tokens.}
We find that 60\% of bucket names contain
at least one corpus token, and 2.3\% contain a l33t spelling variation. The majority of tokens (53\%) are matched to a dictionary, while 14.2\% are common technology names, 10\% symbols, 9.4\% passwords (an existing zxcvbn corpus), 8.4\% human names (an existing zxcvbn corpus), and 4.7\% domain names.    
Corpus tokens are reused, with the top~100 and top~1000 tokens appearing in 25\% and 35\% of all bucket names, respectively (Figure~\ref{fig:token_dist}). The most popular corpus tokens are ``-'' (3.9\%), ``prod'' (0.7\%), and ``test'' (0.6\%). 

However, 40\% of bucket names contain at least one completely random token.
Some of the tokens we label as random may simply be composed of words not found in any of our corpuses. For example, the most common random tokens are \texttt{s} (0.7\%), \texttt{-2} (0.7\%), and \texttt{-us} (0.5\%). We manually investigate the 1,000~most common ``random'' tokens and identify that 10.6\% contain technology-related terms that did not appear in our original corpus and the remaining 89.4\% are indeed random alphanumeric strings.  
Furthermore, more than 99.9\% of random tokens each occur in fewer than 0.00001\% of buckets and 93\% occur in only a single bucket name, which suggests that additional dictionaries have limited utility for predicting the most valid bucket names.


\begin{table}[h]
    \centering
    \small
    \begin{tabular}{lll}
    \toprule
          Name &\% & AVG    \\
           Structure & Names &  $\log_{10}(Guess)$ \\
           \midrule
         (rand) &  28.5\% & 13.7\\
          \quad Length 18 & 24.1\% & 18.0 \\
          \quad Length 25 & 12.4\% & 25.0  \\
          \quad Length 4 & 7.2\% & 4.0 \vspace{2pt} \\
          (rand,corpus) &  12.0\%&8.8\\  
          \quad (rand,``test'') & 3.8\%& 8.1\\  
          \quad (rand,``static'') & 3.1\%& 7.5\\  
          \quad (rand,``img'') & 2.8\%& 7.1 \vspace{2pt}\\  
         (rand,corpus,rand) &  9.8\%&17.5\\
         \quad (rand,``upload'',rand) & 1.2\% & 11.1 \\
         \quad (rand,``asset'',rand) & 1.1\%&11.1  \\
         \quad (rand,``log'',rand) & 0.9\%&20.5 \vspace{2pt}\\
        (corpus,rand) & 6.5\%& 9.5\\  
         \quad (``test'',rand) & 1.9\%& 7.3\\  
         \quad (``img'',rand) & 1.4\%& 8.3\\  
         \quad (``static'',rand) & 1.3\%& 8.5   \vspace{2pt}  \\  
        (corpus,rand,corpus) & 3.3\%&11.7\\
        \quad (``staging'',rand,``appspot.com'') & 0.9\% & 23.1  \\
        \quad (tech term, ``s.'', top domain) & 0.9\% & 9.6 \\
        \quad (tech term,``s.'', dict) & 0.2\% & 8.2 \\
    \bottomrule
    \end{tabular}
    \caption{\textbf{Top 5 Bucket Naming Patterns}---\textnormal{The top~5 naming patterns account for 60\% of buckets. Random strings are the most common naming structures, comprising 28.5\% of all bucket names. }
    }
    \label{tab:naming_schemes_passive}
\end{table}

\vspace{5pt}
\noindent \textbf{Naming Patterns.}
Naming patterns (i.e., specific concatenations of corpus and random tokens produced by zxcvbn) are prevalent across bucket names and are surprisingly predictable. We uncover 3,440~unique naming patterns, but find that the top~5 account for 60\% of buckets and top~60 account for 90\% of buckets. In general, most names are composed of random tokens concatenated with words appearing in one of our corpuses; 92\% of buckets contain at least one random token, 60\% contain at least one known word, and 53\% of buckets contain both.
Only 29\% of all buckets (Table~\ref{tab:naming_schemes_passive}) are a single random character string.


The location of tokens in a naming pattern is surprisingly predictable: 68\% of tokens that appear more than once appear in the same location (e.g., the first word in a pattern) over 50\% of the time and 43\% appear in the same location 100\% of the time. For example, the random tokens ``ules-'' and ``zcb\_'' that appear in 1,159 and 755~buckets, respectively, appear as the second and first token, respectively, in a naming pattern 100\% of the time. Random tokens that appear in at least 10~bucket names are ten times more likely to always appear in the same index compared to corpus tokens.



\subsection{Summary}
Existing active scanning methods are biased towards finding ``easy'' names because they (1) do not rely enough on existing corpuses---82\% of bucket names found by Continella are random strings (e.g., do not contain a word from any of our provided corpuses), and (2) are oblivious to the structures of real-world names. 
Consequently, state-of-the-art scanners underestimate the fraction of vulnerable buckets by up to 3.2~times. 
However, bucket name patterns are prevalent---with the top~5 patterns accounting for 60\% of bucket names---and the locations of tokens in naming patterns are stable: 43\% of tokens appear in the same place 100\% of the time.
In the next section, we introduce an intelligent cloud storage scanning system that uses existing bucket names as training data to find bucket names following a more similar distribution to those found in the wild.

\section{\gens: A System for Finding Real-World Buckets}
\label{sec:generators}

In this section we introduce \gens, a system that ``looks down into the clouds,'' by efficiently predicting valid and vulnerable buckets with the same semantic complexity as buckets found in passive data sources. \gens achieves its performance by leveraging existing research in the password space. Using bucket names found in passive DNS sources from Section~\ref{sec:motivation} as training data, \gens is able to find buckets with a 4~times higher long-term hit-rate,  4~orders of magnitudes more complex names, and find 2.4~times more misconfigured buckets than existing solutions. \gens is available at \url{https://github.com/stanford-esrg/stratosphere} under the Apache 2.0 license. We note, and further elaborate in the section, that its design naturally restricts its use to researchers with a privileged viewpoint to train against (e.g., passive DNS data), thereby limiting an attacker's ability to abuse \gens. 

\subsection{Leveraging Password Cracking}
\label{sub:sec:pass_crack}
Cloud storage bucket names encapsulate the process of humans using language to construct strings to protect their data. 
While prior work has not previously investigated human generated bucket names, there has been extensive work studying passwords, which is another example of human-named strings that can be guessed to reveal data~\cite{wash2016understanding,oesch2020then}.
The password space has excelled at using patterns, such as commonly used passwords, to predict (``crack'') human-named strings~\cite{Cracking,CFG,nn}.
We draw on the similarities between bucket naming patterns and password composition and  evaluate three algorithms previously used to successfully guess passwords to guess valid bucket names: Probabilistic Context Free Grammar (PCFG)~\cite{CFG}, N-Grams~\cite{Cracking} and Long Short-Term Memory (LSTM)~\cite{nn}. 

We describe below how we adopt each password generator to predict bucket naming patterns. We apply both the PCFG and N-Grams algorithms to characters and tokens, creating a total of five distinct bucket-name generators.

\vspace{5pt}
\noindent \textbf{Character-Level 5-Grams.}
Consecutive sequences of characters are not randomly distributed. For example, 60\% of bucket names contain at least one English word. Character-level 5-grams uses prior correlations of consecutive characters to predict new successive tokens: $P(X_{i+1}|X_i)$ where $X_i$ is the $i$th character over all consecutive character pairs. Similar to how prior work has used character-level n-grams for fast dictionary attacks on passwords~\cite{narayanan}, we create a distribution of bucket-name lengths from our ground truth set and independently sample a candidate bucket length to determine how many successive characters to generate sequences of a particular length. We then continuously generate characters until we have reached the given length. We choose $N=5$ to balance performance and memory usage~\cite{omen}.


\vspace{5pt}
\noindent \textbf{Token-Level Bigrams.}  
Tokens co-occur across bucket names, with 9.9\% occurring in at least two bucket names. Token-level bigrams extends character-level n-grams to generate sequences with distributions over consecutive tokens instead of over consecutive characters. We split a bucket name into tokens by delimiting on ``\_'', ``-'', and ``.'' characters, since 39\% of buckets from extracted sources contain at least one delimiter token and symbol-delimited tokens follow a distribution similar to zxcvbn (Figure \ref{fig:token_dist}). We construct a distribution of $P(X_{i+1}|X_i)$ where $X_i$ is the $i$th token over all consecutive token pairs. We find $N=2$ to achieve the best token-expressiveness and memory footprint. To generate sequences of a particular length, we create a distribution of token lengths from our ground truth set and independently sample a candidate bucket token length to determine how many successive tokens to generate. We insert a sampled non-alphanumeric delimiter between each consecutive token pair.

\vspace{5pt}
\noindent \textbf{Probabilistic Context Free Grammar (Character PCFG).} 
We use existing naming patterns (Table~\ref{tab:naming_schemes_passive}) to help predict new concatenations of random tokens (as 92\% of bucket names contain at least one random token, and 40\% contain only random tokens). 
A PCFG is a context free grammar that represents grammatical terminals as a distribution of strings. We leverage techniques used by Weir et. al in the password-cracking literature~\cite{CFG} to build a distribution of bucket names to templates of tokens, where consecutive alphabetic characters are grouped by length, consecutive numeric characters are grouped by length, and remaining non-alphanumeric characters are left hard-coded. For example, the bucket name \texttt{name1234-word-4-} is represented as a rule of the form \texttt{C4N4-C4-N1-}. We then guess bucket names by first generating a bucket rule according to a global distribution of bucket rules and then generate each contiguous alphanumeric token according to a local distribution of tokens that conform to that terminal variable (e.g., C4).

\vspace{5pt}
\noindent \textbf{Token PCFG.}
We use existing naming patterns (Table~\ref{tab:naming_schemes_passive}) to help predict new concatenations of existing tokens (as the top~1000 tokens appear in 35\% of all bucket names). We modify the Character PCFG to create templates of tokens using the patterns in Section~\ref{sec:motivation}. For example, the bucket name ``name1234-word-jpg-4-'' is represented as a rule of the form ``<other>-<dictionary word>-<file exension>-<number>-.'' The token types ``<other>'', ``<dictionary word>'', and ``<file extension>'' will store a frequency list of each seen token of that type. Each rule is also stored in a frequency list. To generate a bucket, we then generate a rule at random according to its frequency and for each token type sample a token of that type according to its frequency.

\vspace{5pt}
\noindent \textbf{LSTM (RNN).} 
A Long Short-Term Memory (LSTM) recurrent neural network (RNN) is a neural network that generates sequences of predicted elements and is able to find hidden patterns that might have not been explored in Section~\ref{sec:motivation}. RNNs have been used to generate password sequences in prior work~\cite{nn}. We use the LSTM to generate sequences of text that are similar to our ground-truth set of buckets.  The LSTM, given a sequence of characters in the format described above, outputs a probability distribution of the following character. Our generator then draws a character from this multinomial distribution, appends the character to the current candidate name, and repeats the generation process. Finally, when a termination character is predicted, the generator submits this candidate name and checks that the name has not been validated before (Section~\ref{sec:motivation}). This model implicitly encodes more state than a Character Grams generator because it considers a full sequence beyond just the previous token when making predictions. 
We leverage Keras~\cite{keras}, a library on top of TensorFlow, for building and training the LSTM.

\subsection{\gens Design}

\begin{figure}[h]
\includegraphics[scale=0.45]{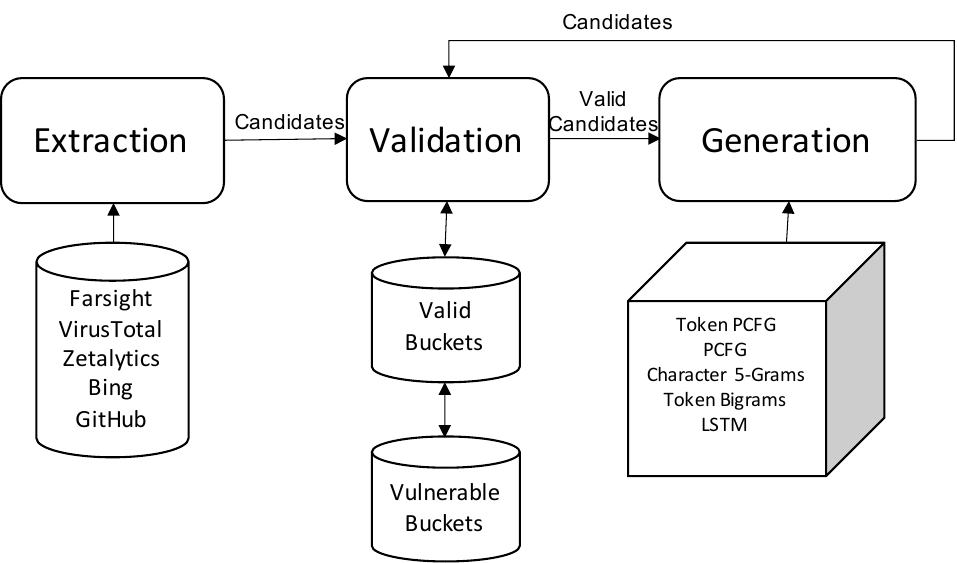}
\caption{\textbf{\gens Design}---\textnormal{\gens consists of a three-stage pipeline that allows it to use existing found buckets to predict new buckets with similar semantic structure.}}
\end{figure}

\gens is composed of a three-stage pipeline: extraction, validation, and generation. During extraction, all names observed in our passive DNS and search sources (Section~\ref{sub:sec:gtruth}) are fetched. During validation, all extracted names are checked across the AWS, GCP, and Alibaba clouds using the process described in Section~\ref{sub:sec:gtruth}. During generation, the user chooses a generation algorithm (or ensemble of algorithms) of their choice. The generation algorithm then uses the previously validated names as training data to train itself. Once trained, batches of bucket candidates are generated and the validation and generation process are repeated. 

\gens's performance is highly dependent on the quality of training data. We note that two of our primary passive data sources (Farsight, Virustotal) only provide free access to their data to verified academics. Consequently, the costs are higher for malicious parties to gain access to high-quality training data, thereby limiting an attacker's ability to abuse \gens.

\subsection{Performance Evaluation}


We evaluate five different versions of \gens, each version using a different individual generating algorithm from Section~\ref{sub:sec:pass_crack}. We refer to each version of \gens by its generator name. We evaluate each version of \gens across three metrics: hit-rate, time (to generate valid candidates and train), and complexity of bucket names found.
We discover that Token PCFG achieves the highest hit-rate and finds the most complex bucket names.

\begin{figure}[h]
    \centering
    \includegraphics[width=\linewidth]{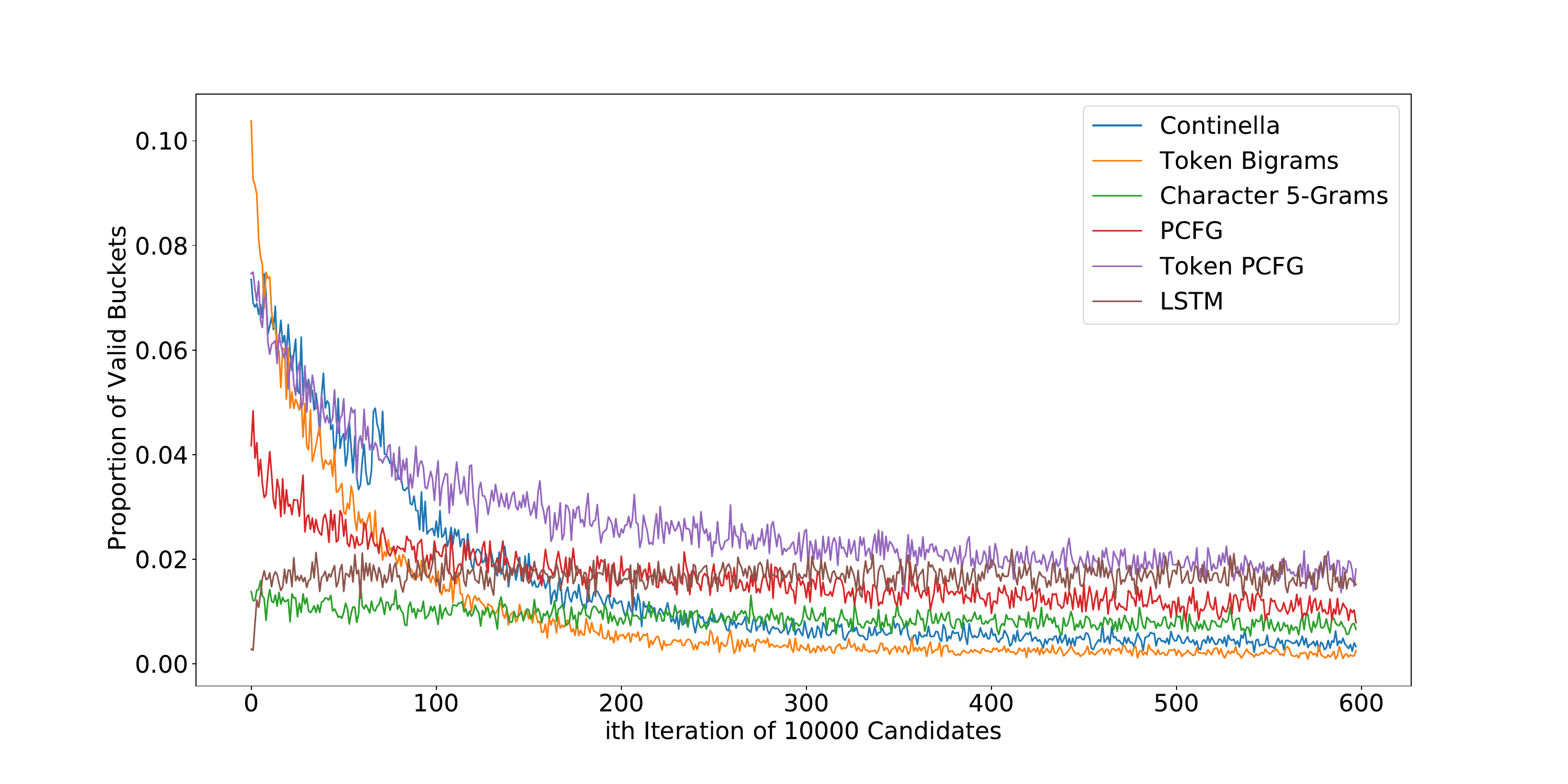}
    \caption{ \textbf{Generator Hit-Rate Over Time}---\textnormal{During the first generation iteration, Token Bigrams finds the most valid buckets (11\%). Over time, all generators' accuracy plateaus below 2.5\% (i.e., fewer than 250 buckets per 10,000).}
    }
    \label{fig:Generator Performance}
\end{figure}

\vspace{3pt}
\noindent \textbf{Hit-rate.}
We first consider how often each generator's guesses are correct in Figure~\ref{fig:Generator Performance}. We note that across all iterations, we do not allow bucket names to repeat, and thus 100\% of buckets collected by our scanners are 100\% unique compared to all passive and search datasources.
During the first generation iteration, our Token Bigrams finds the most valid buckets (11\%). However, the generator's accuracy immediately decays. Within 150~iterations, accuracy across all generators plateaus between 0.5\% (Token Bi-grams) and 2.5\% (Token PCFG). Though the neural network (LSTM) is the most complex scanning algorithm, it does not seem to improve, implying that it learns all latent patterns within the first 10~iterations.

In total, the five generators discover 599,016~buckets, of which 393,598~buckets are found by only one generator across all passive, search, repository, and scanning sources (which we consider a ``unique valid bucket''). Of those, 39,846~buckets (7\%) are publicly accessible. The LSTM and Token PCFG discover the most unique buckets (50\%), while the LSTM and the Character-Level 5-Grams have the most stable asymptotic behavior.  
Compared to Continella, all generators except Token Bigrams experience a hit-rate 1.5 to 4~times larger once 3M candidates are guessed, implying that certain generators will be more successful at finding more buckets overtime. 
Furthermore, Token PCFG finds 1.4~and 2.4~times more sensitive and misconfigured buckets, respectively, than Continella. 

\begin{figure*}[h]
\subfigure[Without Training]{%
  \includegraphics[width=0.5\textwidth]{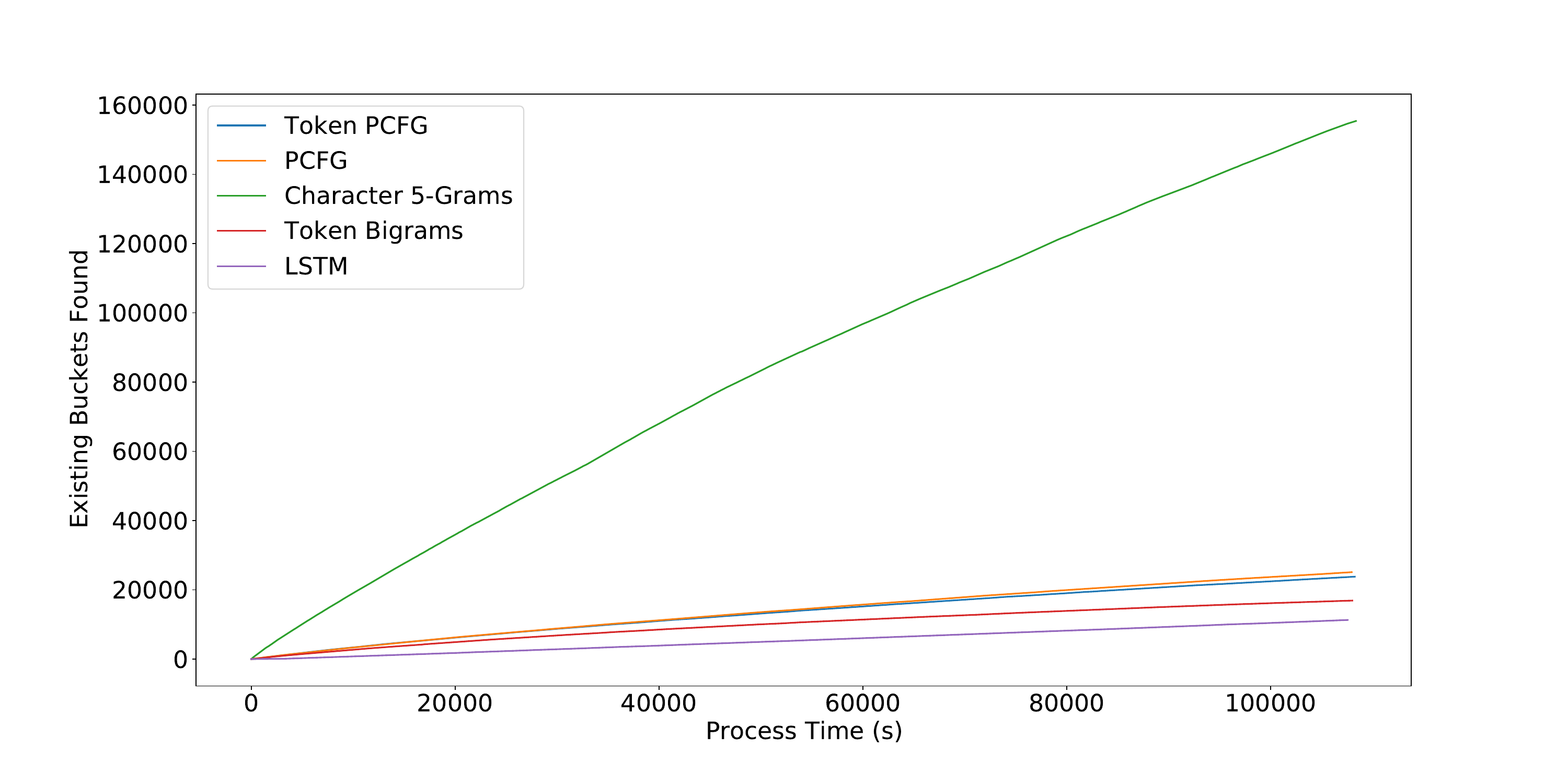}
  \label{fig:generator_time_performance}
  }%
\subfigure[With Training]{%
  \centering
	\includegraphics[width=0.5\textwidth]{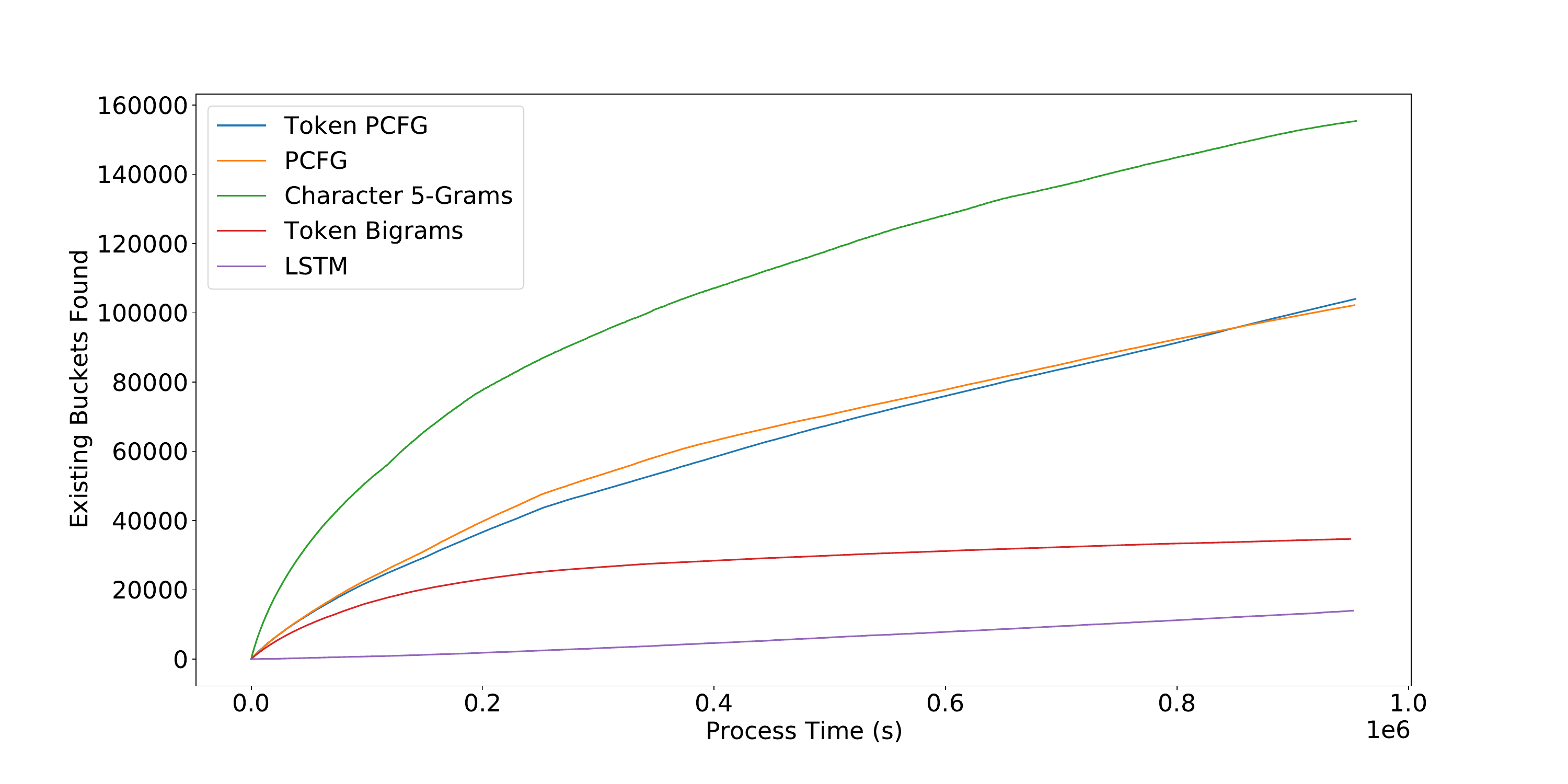}
	\label{fig:generator_train_time_performance}
	 }%
	\caption{\textbf{Generation Time}---%
	\textnormal{ Character 5-Grams is the fastest at generating valid candidates (without training), generating a total of 146,000 valid candidates in 100,000 seconds. Training time adds nearly an order of magnitude more time to produce the same number of valid buckets.  } 
	}
\label{fig:gen_perf}
\end{figure*}

\vspace{3pt}
\noindent \textbf{Time.}
We calculate the core time that each generator takes to generate 10K~valid bucket names over the course of one month (Figure~\ref{fig:generator_time_performance}). Character 5-Grams is the fastest at generating valid candidates, generating a total of 146K valid buckets in the algorithm's first 100,000~seconds. Unsurprisingly, the LSTM---being a neural network---takes the most time to generate a valid candidate, generating only 10,500 valid candidates in the same time period. We further evaluate the process time to generate valid bucket guesses \textit{with} re-training and updating the model every 10K~candidates in Figure~\ref{fig:generator_time_performance}, and find that it adds nearly an order of magnitude more time to generate the same number of candidates (i.e., character 5-grams generates 15.2K valid buckets in 1M seconds). The hit-rate, generation time, and training time provide tradeoffs around how often a generator's model should be updated and the number of guesses each generator should make -- a lower hit-rate but faster generation rate can put greater stress on the network, for example.


\begin{figure*}[h]
\subfigure[Guessability of Generated Candidates]{%
  \includegraphics[width=0.5\textwidth]{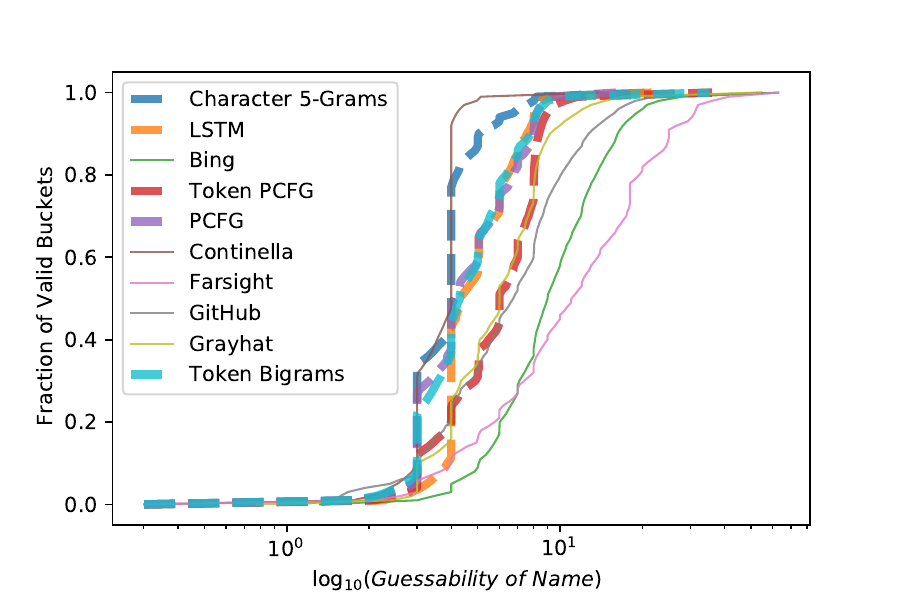}
  \label{fig:gen_guesses}
  }%
\subfigure[Guessability of Generated Candidates Without Random Tokens]{
  \centering
	\includegraphics[width=0.5\textwidth]{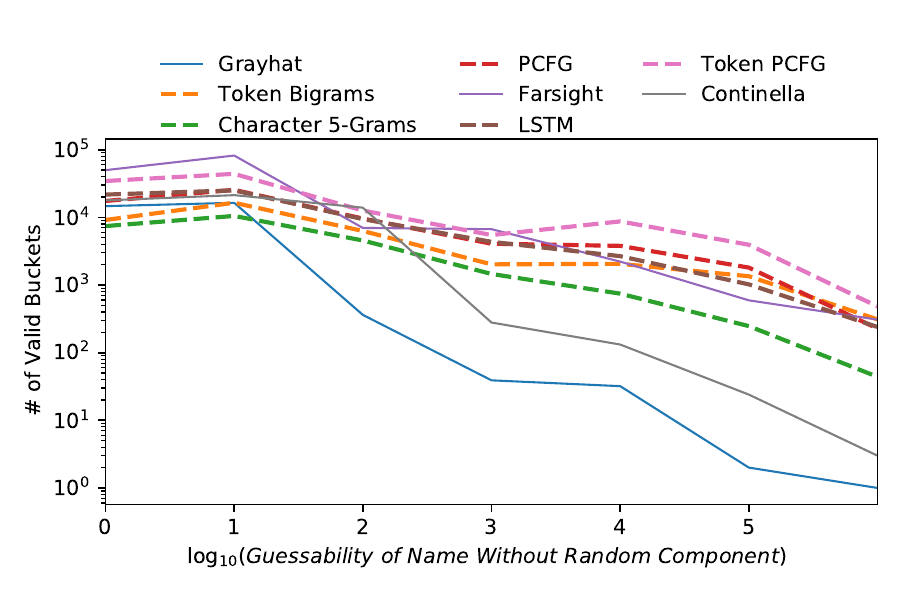}
	\label{fig:gen_guesses_noRand}
	  }
	\caption{\textbf{Guessability of Bucket Names}---%
	\textnormal{(a) The Token PCFG algorithm finds the most complex bucket names (over three magnitudes more complex than Continella). (b) When removing all random tokens in names, all five new scanning approaches find a similar number of complex names compared to Farsight, while existing scanning methods and black-box repositories still fail at guessing the complex structures of names. 
	}}
\label{fig:gen_guesses_b}
\end{figure*}

\vspace{3pt}
\noindent \textbf{Bucket Name Complexity.}
As discussed in Section~\ref{sub:sec:existing_scanners}, while Continella initially appears to have a high success rate at guessing buckets, this is because it guesses short, simple bucket names, effectively focusing on only a small high-density portion of the name space. When running all scanners for one month, all five of the generators find more complex names, with an average and longest length 10~characters and 17~characters longer than Continella, respectively. 
Token PCFG generates the most complex valid bucket names of all generators and existing cloud-storage scanning methods (Figure~\ref{fig:gen_guesses}). It is not surprising that a generator that combines the use of a grammar and corpus is able to find more complex names, as 60\% of names found in passive sources make use of a corpus token.

When our scanners fail to find complex names, it is typically due to their inability to guess the ``random'' component of bucket names; the median and 90th percentile of all random tokens (i.e., the part of a bucket name that was not matched to any corpus according to zxcvbn) found by active scanning is 5 and 15 orders of magnitude less complex than the random tokens found in passive data sources, respectively. To test if failing to guess the random component of a name is the primary reason active scanning methods are unable to find complex names, we replace all random tokens (i.e., Table~\ref{tab:naming_schemes_passive}) with a constant, while keeping the rest of the bucket name intact (i.e., <const>``test''), thereby creating constant-random patterns. Token PCFG is able to find at least one valid bucket-name for 76\% of all constant-random patterns found in the passive data source that occur more than once and 93\% of constant-random patterns that have occurred at least 100~times (i.e., greater than 0.01\% of all found bucket patterns). Continella and Grayhat, in comparison, are able to identify 61\% and 67\% of patterns that occur more than once, respectively.

We also analyze the guessability distribution of constant-random patterns in Figure~\ref{fig:gen_guesses_noRand}, and find that existing work fails to find a significant number of complex constant-random patterns (up to 4~magnitudes less). All five of our generators find nearly as many complex names compared to Farsight (the passive data source with the most complex names) and four of the five algorithms even find up to a magnitude more complex (i.e., guessability greater than $10^{4}$) constant-random patterns than Farsight. These results suggest that, with time, active scanning methods that closely model bucket name patterns (e.g., Token PCFG) are in fact searching for the correct bucket-name patterns, but require more time to accurately guess random components.

\subsection{Limitations of Predicting Bucket Names}
\label{sub:sec:naming_patterns}
To further understand the fundamental limitations of cloud-storage active scanning, such as the ability to find complicated bucket names that are likely to be more vulnerable, we investigate the theoretical limits of active scanning and compare them to \gens with Token PCFG.

Bucket names found in passive DNS and search sources are typically hard to guess randomly: bucket names are a median 13~characters long, the median Shannon entropy is 50~bits, and the median guessability is $10^{11}$ guesses (using zxcvbn's guessability heuristic~\cite{wheeler2016zxcvbn}). Using a dictionary-driven attack with no prior knowledge of patterns to guess a single name with a median-guessability would take approximately 7~hours on a single Intel I5 core~\cite{i5}. Guessing the easiest 50\% of names (Figure~\ref{fig:gt_guesses}) would take 353.4~years, assuming no network overhead.

However, understanding naming patterns can simplify the search space by 9~orders of magnitude, as only 29\% of bucket names are truly random (Table~\ref{tab:naming_schemes_passive}). We recalculate the zxcvbn heuristic assuming that the pattern is known beforehand (i.e., removing the probability of a token occurring in a certain pattern index) and find the median guessability becomes $10^{4}$. In other words, understanding patterns would shorten the time to guess the easiest-guessable 50\% of names to a mere 4~minutes compared to 353.4~years.
Using the hit-rate at which \gens + Token PCFG finds bucket names equivalent to the median guessability of passive sources ($10^{11}$), 0.02\%, it would take \gens + Token PCFG 2.4~minutes, assuming no network overhead, to guess the easiest-guessable 50\% of names found in all of our passive datasources. 

\subsection{Summary}
We present \gens, the first cloud-storage scanning system that uses previously discovered bucket naming patterns to search the entire cloud-storage name space and achieve a long-term higher (up to 4~times) hit-rate compared to existing active scanners. 
Though random components of complex bucket names create a fundamental limitation for active scanning, \gens + Token PCFG, unlike existing scanners, is still able to find 93\% of all popular naming patterns and nearly the same number of hard-to-guess names as Farsight when ignoring random tokens, likely approaching the fundamental limits of what a scanner can achieve. Finding complex names also allows \gens + Token PCFG to find 1.4~and 2.3~times more sensitive and misconfigured buckets than Continella when scanning for an equivalent time. 
\section{Security Analysis of Found Buckets}
\label{sec:security}
We combine all 2.1M~valid cloud-storage buckets found across our eight data sources and five variations of \gens to analyze the security posture of the cloud storage ecosystem. By quantifying the exposure of sensitive files, misconfigured bucket permissions, and the active exploitation of storage buckets, we discover previous work~\cite{Continella} underestimates the vulnerability of cloud storage by up to 5.8~times. 

\subsection{Exposed Sensitive Files}

Of the 2.1M~valid buckets we find, 173K (13\%) have publicly listable files. However, this does not directly indicate a vulnerability---some buckets may intentionally allow public access. For example, buckets used to host public websites or to share public data sets are typically public. To better understand whether public buckets pose a security risk, we investigate the types of files exposed by collecting file metadata for the first 100K~files in each bucket. We use American Express EarlyBird~\cite{earlybird} (Section~\ref{sec:motivation}) to detect whether sensitive data is exposed. In total, 10.6\% of public buckets host sensitive data---5.3~times more than reported in previous work~\cite{Continella}---of which 6\% contain backup files, 4\% contain database dumps, and 3.6\% buckets contain cryptographic keys. 
In aggregate, we find hundreds of thousands of sensitive files:
14.4K~private keys (.p12, .key, or .pfx), 683K~SQL dumps (.sql or .sql.gz), and 99.4K~backups (.bak). 

By manually analyzing filenames, we discover 10~notably sensitive buckets that clearly belong to a private company or organization, including a travel agency publicly revealing over 8K~travel documents (including passports), a popular messaging application's repository of 20K~user uploads, a Point of Sale restaurant management app's collection of production RSA private keys, an identity protection company's customer invoices, and a Department of Defense contractor's internal documentation. 
We further detail the disclosure process of the 10~buckets in Appendix~\ref{app:disclose}. 
    
\subsection{Misconfigured Bucket Permissions}

\begin{table*}[h]
\centering
\small
\begin{tabular}{l l l l l l l }
\toprule
& \multicolumn{2}{c}{Amazon AWS} &  \multicolumn{2}{c}{Google Cloud} & \multicolumn{2}{c}{Alibaba} \\ \cmidrule(r){2-3} \cmidrule(r){4-5} \cmidrule(r){6-7}
   &   \% Public & \% Private & \% Public & \% Private & \% Public & \% Private \\ 

\midrule
%

Read Permissions (Bucket)&23.54\% (25,587)&1.05\% (11,753)&21.37\% (13,539)&25.68\% (114,817)&0.50\% (123)&0.03\% (92)\\

\midrule
List (Objects)&100.00\% (25,587)&0.00\% (0)&100.00\% (13,539)&0.00\% (0)&100.00\% (123)&0.00\% (0)\\
Read (Objects)&73.86\% (18,899)&4.90\% (576)&76.02\% (10,292)&1.01\% (1,157)&100.00\% (123)&0.00\% (0)\\
Write (Objects)&13.40\% (3,429)&4.82\% (566)&6.04\% (818)&0.06\% (66)&56.91\% (70)&0.00\% (0)\\
Change Permissions (Bucket)&8.42\% (2,154)&5.41\% (636)&3.46\% (469)&0.00\% (0)&0.00\% (0)&0.00\% (0)\\
Delete (Objects)&13.40\% (3,429)&4.82\% (566)&5.35\% (725)&0.00\% (0)&56.91\% (70)&0.00\% (0)\\

\bottomrule
\end{tabular}

\caption{ \textbf{Vulnerable Permission Configurations}---\textnormal{Buckets with readable permissions in AWS, compared to GCP and Alibaba, are most likely to contain vulnerable permissions (i.e., buckets that allow writing new content, deleting files, or changing the bucket's ACLs). Most concerning, 5.41\% of private AWS buckets with readable permissions are configured to allow anyone to change the bucket ACL. (Note: Read permission percentages are out of the total number of public and private buckets, while other permission percentages are only out of buckets that allow read permissions.)}}
\label{table:s3_acl}
\label{table:guess_acl}
\end{table*}

\begin{figure}[h!]
    \centering
    \includegraphics[width=\linewidth]{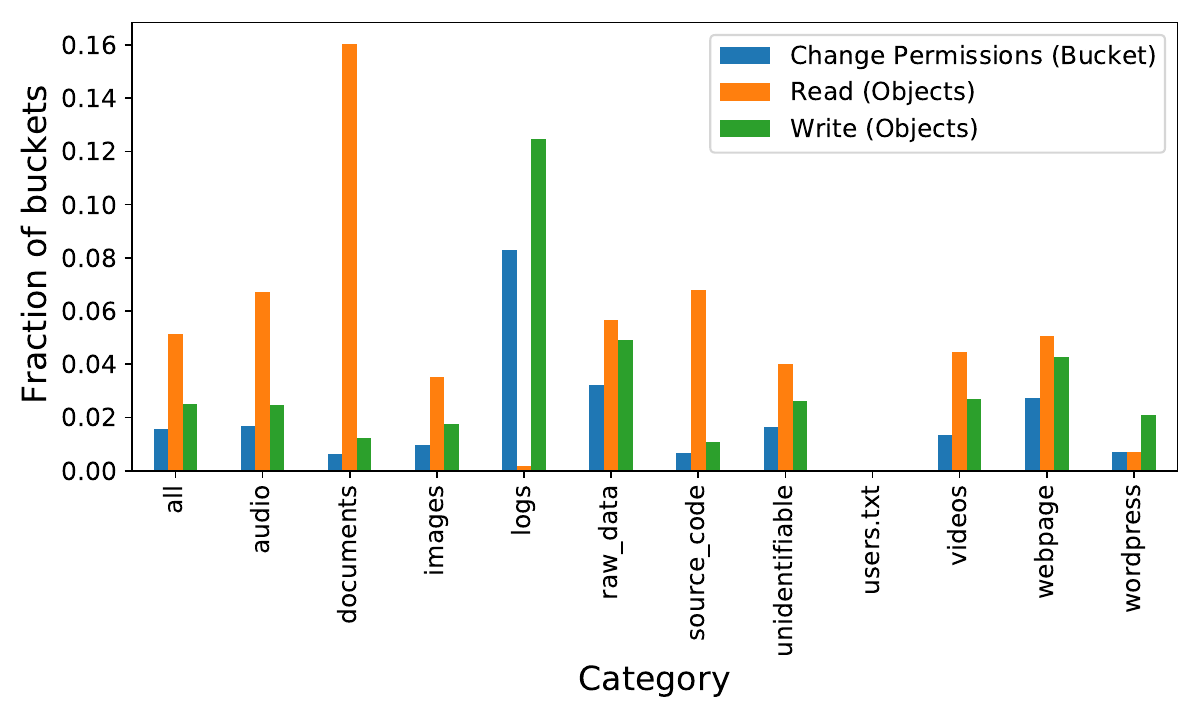}
    \caption{\textbf{Misconfiguration rates by bucket category}---\textnormal{Buckets storing primarily logs are the most likely to be publicly-writable, while buckets containing documents are the most likely to allow reading objects.}} 
    \label{fig:bucket_cat_misconfigs}
\end{figure}

Beyond simply allowing public access, buckets also frequently have vulnerable permissions that allow attackers to delete or read files in what would otherwise be a private bucket. We analyze all publicly available ACLs (23.54\%, 21.37\%, and 0.50\% of AWS, GCP, and Alibaba's public buckets, respectively) and find 13\%, 6\%, and 57\% of AWS, GCP, and Alibaba buckets allow writing new content or deleting existing files (Table~\ref{table:s3_acl}).
Further, 5\% of private AWS buckets with readable permissions (636~buckets) allow for permissions to be changed by the public, thus making their private status irrelevant. Such unintentional exposure of files is only found in AWS, and is likely due to AWS' offering of over 100~unique actions/permissions~\cite{awsPerm} compared to GCP's 14~\cite{gcpPerm}. Alibaba's ACL permission model prevents the problem by allowing only 3~permission states: private, public-read, public-read-write~\cite{alibabaacl}. We show the categories of buckets, using a categorization method detailed in Appendix~\ref{app:filter}, most often misconfigured in Figure~\ref{fig:bucket_cat_misconfigs}. Overall, previous work~\cite{Continella} under-estimated by 5.8~times the number of misconfigured buckets. 

\begin{figure}[h]
    \centering
    \includegraphics[width=\linewidth]{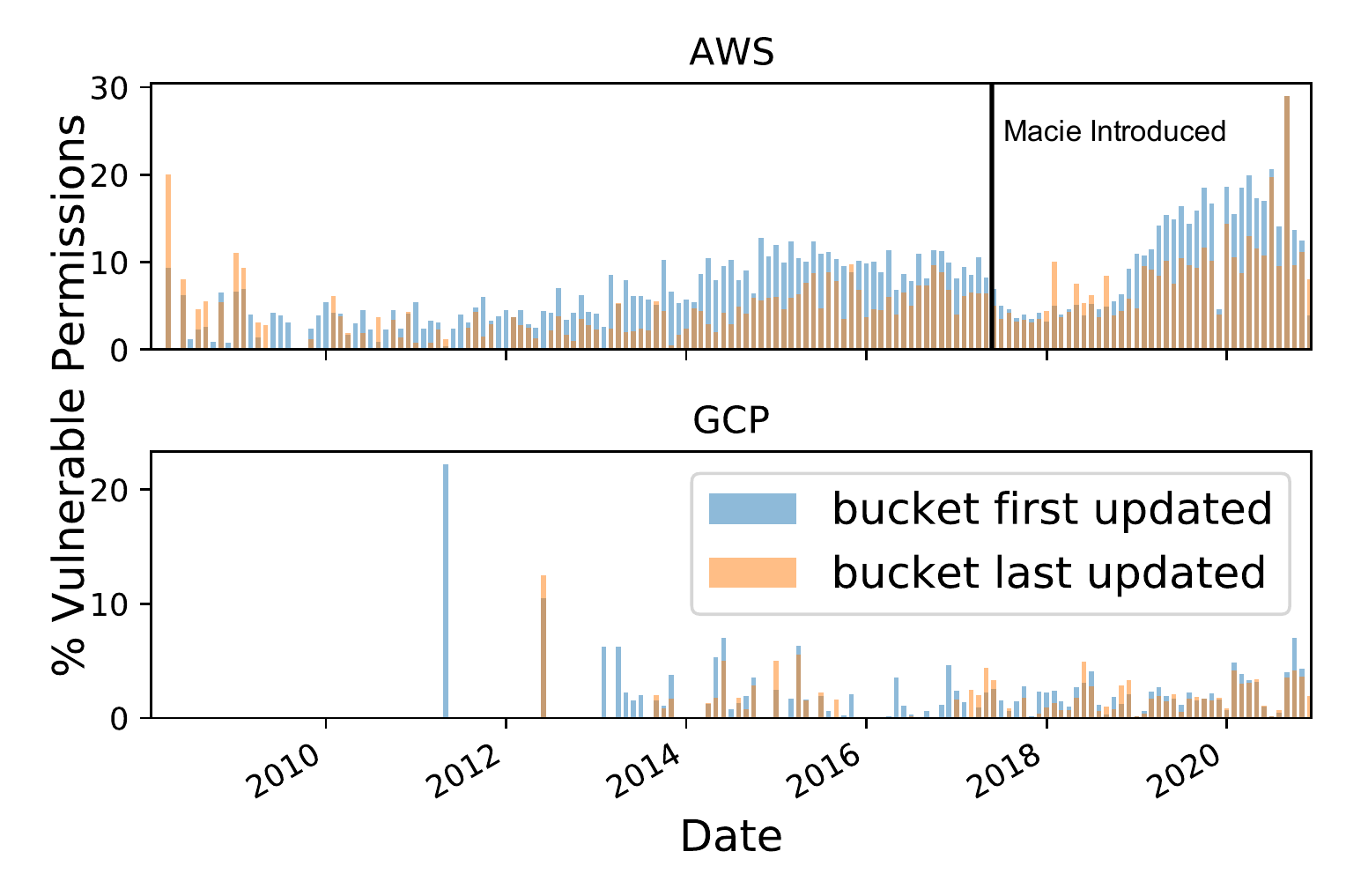}
    \caption{\textbf{Vulnerability of buckets over time}---\textnormal{The fraction of buckets created or last updated on AWS within the past year are on average 4x more likely to be vulnerable compared to buckets created or updated 10~years ago. However, no such pattern exists on GCP.}}
    \label{fig:bucket_vuln_time}
\end{figure}

Vulnerable permissions are a worsening problem for AWS---the fraction of vulnerable public buckets (i.e., buckets that allow writing new content, deleting files, or changing the bucket's ACLs) in our study increases over time for Amazon while remaining relatively stable for GCP (Figure~\ref{fig:bucket_vuln_time}). 
In AWS, buckets that were first or last updated within the past year are on average 1.6~and 4~times more likely to be vulnerable compared to buckets created or updated 5 and 10~years ago, respectively. This may be in part to Amazon's permission scheme becoming more complex over that time period~\cite{s3permissionstight}.


We notice a dip in vulnerable buckets in 2017--2018 that coincides with the launch of Amazon Macie, a service that applies machine learning to detect sensitive information exposed in public buckets~\cite{macie} and a cluster of S3 security articles written on AWS's blog~\cite{s3sec1, s3sec2}. 
Nonetheless, within a year the fraction of vulnerable buckets begins to rise again and become more prevalent in 2020, despite AWS's launch of other S3 security features in late 2018~\cite{blocks,guarduty}.
We note that many factors likely contribute to the apparent increase in vulnerability, including the decreased barrier of entry to using cloud services.


\subsection{Exploited Buckets in the Wild}
\label{sub:sec:warned}

Buckets with misconfigured permissions suffer active exploitation. 
To estimate the fraction of buckets that have been unsolicitedly written-to, we look for evidence of scanning campaigns that upload identical files across buckets. Concretely, we manually explore the top 100 file hashes (appearing in 45K buckets), using ETags (a unique and deterministic file identifier that is often simply the MD5 digest of file contents~\cite{s3ETag, kaminsky2005md5}) that appear as a popular Google search result.
We discover six unique messages left by scanners intended to alert owners of their bucket being publicly writable (i.e., ``white-hat messages'') and six unique messages containing generic content (e.g. empty file, the word "test"). 
Message uploads are often clustered around specific dates---the top 5 days of most warning messages uploaded across 10 years make up 54.6\% of all warning messages---suggesting that messages are left as a result of bulk scans.
We observe a total of 7,475 warning messages uploaded to 4,769 public buckets, with 99.8\% belonging to AWS S3. In other words, 3\%  of all public buckets have already had their vulnerability to being publicly writable exploited.



Surprisingly, 46\% of bucket owners do not notice the exploitation of their bucket and
continue to upload files (ranging from 1~week to 8~years later).
To compare the effects of ``warned'' and ``not-warned'' buckets, we scan all public buckets found one month after their initial scan and conduct a hypothesis test~\cite{kohavi2007practical}; we discover that buckets that were warned in 2020 are statistically significantly ($p<0.05$) more likely  (1.3~times) to turn private or be deleted (``patched'') than  public buckets that were last updated in 2020.
However,  buckets warned in 2019 or years prior are statistically significantly less likely (up to 7~times) to turn private or be deleted than non-warned bucket. 

To further analyze bucket exploitation, we deploy bucket honeypots on the AWS, GCP, and Alibaba clouds (Appendix~\ref{app:honey_meths}); buckets see unsolicited traffic within the first 24~hours of being created. 
Notably, only four months into our study do we receive a warning---only from AWS---that our honeypots are publicly accessible.
From October 2020 to January 2021, we receive a total of 563 requests to our honeypot buckets, with AWS buckets seeing over 4x~the amount of traffic compared to GCP and Alibaba.
The distribution of source traffic is different than the breakdown of IPv4 web scanning~\cite{durumeric2014scanners}: most notably, China leads large IPv4 scanning campaigns but is completely absent from scanning serverless storage. Further, 11\% of bucket requests originate from known Tor exit nodes.

Despite our honeypot buckets being writeable, no sources attempt to write to any of the buckets. Nonetheless, many sources do attempt to enumerate permissions, sending a total of 47 and 5 requests to the AWS and GCP buckets, respectively. Based on request user agents, we observe two instances in which sources attempt to manually view open accessible files after discovering a public bucket.

\subsection{Summary}

Across the three cloud providers, we discover the most popular cloud, AWS, to be the most vulnerable to insecure user configurations, and its vulnerability to be 5~times worse compared to previous work~\cite{Continella}. Moreover, buckets with vulnerable permissions are only becoming more prevalent on AWS, with buckets created in 2020 being 4~times more likely to be vulnerable compared to buckets created or updated 10 years ago. We see that security interventions (e.g., AWS Macie and warning buckets) do decrease the fraction of vulnerable buckets, but only in the short-term.


\section{Discussion and Conclusion}

Misconfigured storage buckets continue to cause catastrophic data leaks for companies. Our results indicate that attackers are actively scanning for vulnerable buckets and that existing defensive solutions are insufficient for uncovering problems before attackers do. We show that while state-of-the-art solutions appear to initially have high hit-rates for guessing buckets, they do not effectively find the types of buckets that are misconfigured or contain sensitive data. Vulnerable buckets tend to have more complex names than previous scanners could uncover. 
Despite insecure buckets having more complex names, many buckets are composed of multiple tokens, incorporate common English words, and are ``human-readable'', which allows us to more efficiently predicting the names of buckets. We show that ML-driven approaches can uncover a significant number of buckets, but that all solutions are fundamentally limited by the frequent inclusion of opaque, random identifiers that are difficult to guess by any algorithm.
Nonetheless, we stress that cloud storage security should not rely on hard-to-guess names, but rather rely on operators securely configuring bucket permissions.

When considering a more comprehensive set of buckets, we find that insecure buckets are much more common than previously believed. Our perspective also helps identify that Amazon S3 buckets have considerably worse security than other providers. This is very likely due to Amazon's particularly complex permissions model. We encourage Amazon to consider either simplifying their permissions model or providing better tools to assist users in uncovering security problems and inconsistent configurations. 
We hope \gens will help researchers more accurately and comprehensively understand and improve the cloud storage ecosystem but emphasize that 
cloud providers need to step up in providing more effective means for keeping bucket contents secure.

\begin{acks}
The authors thank Deepak Kumar, Kimberly Ruth, Katherine Izhikevich, Tatyana Izhikevich and the Stanford University Empirical Security Research Group for providing insightful discussion and comments on various versions
of this work. We also thank VirusTotal, FarSight, and Zetalytics for providing academic access to their APIs.
This work was supported in part by Google., Inc., NSF Graduate Fellowship
DGE-1656518, and a Stanford Graduate Fellowship.
\end{acks}

{\footnotesize \balance \bibliographystyle{ACM-Reference-Format}
\bibliography{reference}}
\pagebreak
\appendix

\setcounter{secnumdepth}{0}
\section{Appendix}

 \setcounter{secnumdepth}{1}
 \section{Honeypot Methodology}
 \label{app:honey_meths}

We deploy storage honeypots on AWS, GCP, and Alibaba, and we find that adversaries actively scan for vulnerable storage buckets. 
We create 35~identically named buckets on each of the three cloud platforms in October, 2020. Of these, five bucket names are English dictionary words, five are organization and adjective combinations (e.g. "Stanford-production"), five are 4, 5, and 6-character randomly generated alphanumeric strings, and ten are 16-character randomly generated alphanumeric strings. We do not publish these bucket names with the exception of five 16-character bucket names, which we commit to a public GitHub repository. We configure all buckets to be publicly readable and writeable. 

\setcounter{secnumdepth}{2}
\section{Coordinated Disclosure}
\label{app:disclose}

Over the course of this research, we encountered 10 cases where we confirmed likely sensitive information was being exposed in public S3 buckets. Our disclosure attempts are detailed as follows. In aggregate we received a response from 6 of 10 disclosures, and 5 of 10 buckets disclosed are still public.

\vspace{2pt}
\noindent
\textbf{Organization 1.} A defense contractor exposed hundreds of document names in a publicly listable S3 bucket. While the files were not accessible, the file naming and metadata revealed considerable information about the company's operations. We disclosed to the company via their vulnerability disclosure policy on 12/03/2019 and have not received a response, and the bucket is still public.

\vspace{2pt}
\noindent
\textbf{Organization 2.} A Point of Sale restaurant management app exposed production RSA keys for their customers. We disclosed to the company via a contact email address on 12/03/2019 and have not received a response, though access is now restricted.

\vspace{2pt}
\noindent
\textbf{Organization 3.} An identity protection company exposed customer invoices. We disclosed to the company via a contact email address on 12/03/2019 and have not received a response, and the bucket is still public.

\vspace{2pt}
\noindent
\textbf{Organization 4.} A software conference exposed thousands of travel documents of aid recipients. We disclosed to the organization via a contact email address on 12/05/2019 and received a response the same day, and access is now restricted. This triggered an audit within the organization to search for similar exposures.

\vspace{2pt}
\noindent
\textbf{Organization 5.} A travel agency exposed over 8,000 travel documents, including passports. As we were unable to determine the owner, we contacted AWS on 12/05/2019, who alerted the owner of the exposure. Access is now restricted.

\vspace{2pt}
\noindent
\textbf{Organization 6.} A startup exposed hundreds of customer invoices. We disclosed to the company via a security contact email address on 12/08/2019 and received a response the next day, and access is now restricted.

\vspace{2pt}
\noindent
\textbf{Organization 7.} A property management company exposed thousands of internal company documents. We disclosed to the company via a contact email address on 12/08/2019 and received a response several days later that the company is aware of the issue, though the bucket is still public.

\vspace{2pt}
\noindent
\textbf{Organization 8.} A computer repair company exposed thousands of customer receipts. We disclosed to the company via a contact email address on 12/08/2019 and have not received a response, and the bucket is still public.

\vspace{2pt}
\noindent
\textbf{Organization 9.} A career management startup exposed hundreds of customer receipts and resumes. We disclosed to the company via a contact email address on 12/05/2019 and received a response several days later that the issue was resolved. Though the company disabled public listing of the bucket, we observed that HEAD requests could still be made to objects. We responded with this information and have not received a response, and objects are still accessible.

\vspace{2pt}
\noindent
\textbf{Organization 10.} Over 20,000~personal documents uploaded on a major secure messaging app were exposed by an unknown entity. We reported this on 12/08/2019 to the bug bounty program of the company that operates the messaging app and received a response that the bucket was not operated by the company. More than a year after reporting, the bucket was made private.

\setcounter{secnumdepth}{3}
\section{Bucket Categorization}
\label{app:filter}

The following table displays the various keywords used to categorized buckets.
We discover a bi-modal distribution (with a 98\% inflection point) of homogeneity of file-category types within one bucket. Thus, we consider a bucket to belong to a category if at least 98\% of its files match at least one keyword for the category, such as common file names or extensions. 

{\centering
\begin{table}[hbt!]
    \centering
    \begin{tabular}{l p{5cm} }
    \toprule
        Category & Keywords \\
        \midrule
        webpage &js, css, html, woff, assets, less, font-awesome, map, scss, static, main, jquery, robots, ckeditor, plugins, fonts, ttf, vendor, bundle, node\_modules \\
wordpress & wp-content, wp-includes, wp-admin, wordpress \\
images & image, images, png, jpg, jpeg, webp, svg, gif \\
videos & video, videos, mp4, avi, mov, wmv, flv  \\
audio & mp3, wav, m4a, flac, wma, aac \\
logs & log, logging, logs, papertrail, elasticloadbalancing, AWSLogs, access-logs, access-logs-v2 \\
documents & pdf, doc, docx, reports, documents, invoices, attachments, invoice, uploads \\
raw\_data & csv, sql, json, xml, gz, zip, export, sftp, ftp, exports, db, reports \\
source\_code &php, asp, aspx, cpp, js, go, php3, rb, py, java, cs \\
users.txt & Includes only users.txt file \\    
    \bottomrule
    \end{tabular}
\end{table}
\label{tab:bucket_categories}
}

\end{document}